# InSb Nanowires with Built-In $Ga_xIn_{1-x}Sb$ Tunnel Barriers for Majorana Devices

*Diana Car,[1,‡] Sonia Conesa-Boj,[2,‡] Hao Zhang,[2,‡] Roy L. M. Op het Veld,[2] Michiel W. A. de Moor,[2] Elham M. T. Fadaly,[2] Önder Gül,[2] Sebastian Kölling,[1] Sebastien R. Plissard,[3] Vigdis Toresen,[2] Michael T. Wimmer,[2] Kenji Watanabe,[4] Takashi Taniguchi,[4] Leo P. Kouwenhoven,[2] Erik P. A. M. Bakkers[*,1,2]*

[1] Department of Applied Physics, Eindhoven University of Technology, P.O. Box 513, 5600 MB Eindhoven, the Netherlands

[2] Kavli Institute of Nanoscience, Delft University of Technology, 2628CJ Delft, the Netherlands

[3] CNRS-LAAS, 7 avenue du Colonel Roche, F-31400 Toulouse, France

[4] Advanced Materials Laboratory, National Institute for Materials Science, 1-1 Namiki, Tsukuba 305-0044, Japan

[‡] These authors contributed equally.

*Corresponding author: e.p.a.m.bakkers@tue.nl

**ABSTRACT**. Majorana zero modes (MZMs), prime candidates for topological quantum bits, are detected as zero bias conductance peaks (ZBPs) in tunneling spectroscopy measurements. Implementation of a narrow and high tunnel barrier in the next generation of Majorana devices can help to achieve the theoretically predicted quantized height of the ZBP. We propose a material-oriented approach to engineer a sharp and narrow tunnel barrier by synthesizing a thin axial segment of $Ga_xIn_{1-x}Sb$ within an InSb nanowire. By varying the precursor molar fraction and the growth time, we accurately control the composition and the length of the barriers. The height and the width of the $Ga_xIn_{1-x}Sb$ tunnel barrier are extracted from the Wentzel-Kramers-Brillouin (WKB)-fits to the experimental I-V traces.

Semiconductor heterostructures have revolutionized solid-state physics by providing the opportunity to manipulate the motion of electrons and holes via band engineering. Physical properties of heterostructures are for an important part defined by the interfaces, which therefore need to be of highest quality, *i.e.* free of any structural defects. For this reason, only a few, lattice matched III-V materials are commonly used in heterostructures, such as GaAs, AlAs and the intermediate alloys AlGaAs;[1] AlInAs, GaInAs and InP;[2] GaAs, InGaP and AlInP[3] etc.

InSb is attractive for high-speed, low-power electronics,[4,5] infrared optoelectronics,[6] thermoelectric power generation[7] as well as spintronics[8,9] and topological quantum computing[10] due to the highest electron mobility and the narrowest (direct) bandgap of all the III-V semiconductors, as well as a large Landé *g* factor[11] and strong spin-orbit interactions.[12] To profit further from the advantageous properties of InSb it is of crucial importance to integrate this semiconductor in high-quality heterostructures.

Epitaxy of InSb-based heterostructures is difficult due to the large lattice parameter of InSb: the lattice mismatch between InSb and its nearest III-V neighbor GaSb is ~ 6.3%. However, due to their nanoscale diameter and high aspect ratio, nanowires allow for stacking of lattice mismatched materials which would be impossible to realize in planar geometries.[13] Nevertheless, synthesis of axial nanowire heterostructures is not an easy task. The use of a metal seed particle introduces complications not present in thin-film systems, such as the difficulty to form sharp interfaces due to the reservoir effect.[14,15] Since the solubility of the group III species in the catalyst particle is much higher than the solubility of the group V species, it is especially challenging to form sharp interfaces when switching the group III materials. Indeed, atomically sharp interfaces have been reported in heterostructures formed by group V switching,[16–20] while heterostructures grown by switching of the group III material normally show graded interfaces.[21–26] In addition, nanowire heterostructures grown by switching of the group III species often show kinking,[27] undesired radial growth,[25] shift of the Au droplet during growth[21,28] or diameter-modulation.[29,30]

Here, the growth, structural and electronic properties of InSb/$Ga_xIn_{1-x}Sb$/InSb nanowire heterostructures are reported. $Ga_xIn_{1-x}Sb$ has a larger bandgap than InSb, and a type-I band alignment is expected for this system in which the $Ga_xIn_{1-x}Sb$ segment acts as a barrier for both electrons and holes.[31]

One important motivation for this specific material combination is that one dimensional InSb/$Ga_xIn_{1-x}Sb$/InSb axial heterostructures provide a suitable testing ground for the new functionalities of quantum mechanical devices, such as Cooper-pair splitters[32,33] and hybrid superconductor-semiconducting nanowire devices used for detection of Majorana zero modes (MZMs).[34–38] Despite the significant improvements in the performance of Majorana-devices that have been reported recently,[39,40] the height of the ZBP remains much lower (~$0.1G_0$[39]) than the predicted, quantized ZBP height of $G_0=2e^2/h$ at 0 temperature.[41] Several theoretical works[42–45] have pinpointed the smooth tunnel barrier as a possible cause of the weak ZBP. Measuring a quantized ZBP is important because it would be a direct evidence of the topological nature of this phenomenon.[46]

In this work, we achieve sharp interfaces, high structural quality and full control over the Ga fraction content and the width of the barriers.

InSb/$Ga_xIn_{1-x}Sb$/InSb nanowire heterostructures have been synthesized by a Au-catalyzed Vapor-Liquid-Solid (VLS) growth mechanism in an Aixtron Metal Organic Vapor Phase Epitaxy (MOVPE) machine using tri-methyl-indium (TMI), tri-ethyl-gallium (TEGa) and tri-methyl-antimony (TMSb) as growth precursors. To investigate the influence of the TEGa molar fraction, $X_i$ (TEGa), on the amount of Gallium incorporated in the barrier, we have grown a set of samples by keeping the growth time fixed to 30 seconds and increasing the molar fraction of TEGa (for detailed explanation of the growth process see **Supporting Information (SI)-1**). The crystal structure and chemical composition of the nanowires were characterized by high-resolution transmission electron microscopy (HRTEM), scanning transmission electron microscopy high-angle annular dark-field imaging (STEM−HAADF), and energy-dispersive X-ray spectroscopy (STEM−EDX).

**Figure 1a** shows a STEM-HAADF image of a representative nanowire where the position of the barrier can be identified as a clear contrast difference across the diameter (indicated by an arrow). From the HRTEM image (**Figure 1b**) performed on the same nanowire we can also identify the thin barrier by a difference in contrast. The associated fast Fourier transform (FFT) (**Figure 1c)** reveals that the crystalline phase is pure zinc blende. We note that all InSb/$Ga_xIn_{1-x}Sb$/InSb nanowires analyzed have pure defect-free zinc blende crystal structure.

From the EDX line scans shown in **Figure 1d-f** we can see that the length of the $Ga_xIn_{1-x}Sb$ barriers of all the 3 samples is 20 nm while the Ga at. % concentration increases from 5% for sample (d) to 14% for sample (f).

The InSb/$Ga_xIn_{1-x}Sb$/InSb nanowires reported here have several favorable properties for device fabrication and transport experiments. The high aspect ratio of the InSb/$Ga_xIn_{1-x}Sb$/InSb nanowires allows for increased freedom in device design. Moreover, InSb/$Ga_xIn_{1-x}Sb$/InSb nanowires are uniform in thickness, *i.e.* they are not tapered and there is no diameter modulation induced by the barrier segment (see **SI-2**). It should be noted that there is no undesired radial growth present: the barrier spans the whole diameter of the nanowire and there is no $Ga_xIn_{1-x}Sb$ shell grown around the InSb segment (see **SI-3**).

To investigate the strain in the barriers, we employ geometrical phase analysis (GPA).[47–50] **Figure 2a** shows a HRTEM image of a nanowire containing a 20 nm wide $Ga_{0.28}In_{0.72}Sb$ barrier. **Figure 2b** shows a higher magnification HRTEM image of the region indicated by a red square in **Figure 2a**. We have selected the (1-11) Bragg reflection to be filtered and analyzed in order to study the strain component $\varepsilon_{zz}$ along the growth direction. InSb region has been chosen as a reference. The resulting strain map is displayed in **Figure 2c**. From the strain profile integrated along the growth direction (**Figure 2d**) the barrier is compressively strained along the *z* - direction with respect to the InSb reference region. Since the lattice parameter of InSb is larger than the lattice parameter of GaSb, we expect the $Ga_{0.28}In_{0.72}Sb$ segment to be tensile strained in the interface plane and compressively strained out-of-plane, *i.e.* along the [1-11] direction. The average value of the measured compressive strain in the $Ga_{0.28}In_{0.72}Sb$ segment is around -2%. A similar analysis for a 20 nm wide $Ga_{0.15}In_{0.85}Sb$ barrier is shown in **SI-4**. Importantly, for both samples analyzed, we do not observe any misfit dislocations induced in the barriers, indicating that the strain is not plastically relaxed.

In order to perform transport measurements, InSb/$Ga_{0.28}In_{0.72}Sb$/InSb nanowires have been transferred to a $SiO_2$-covered p-doped Si-substrate patterned with a set of local metallic gates on top of which a sheet of hexagonal boron nitride (hBN) is mechanically transferred as the dielectric. Using TEM, we have determined the axial position of the barrier (see **SI-5**) for a number of nanowires. The position of the barriers depends on the total length of the InSb wire. A micromanipulator mounted in the chamber of a scanning electron microscope (SEM) is used to deterministically position the barrier segment perpendicularly above one of the fine gates. The ability to control the position of the barrier is of particular importance for applications in hybrid superconductor-semiconducting nanowire devices, since the barrier needs to be precisely aligned with the superconducting contact and the local gates. The Cr/Au ohmic contacts are defined using electron beam lithography. The samples are cooled down to a temperature of 2 K. A detailed description of the fabrication steps can be found in **SI-6**. From the total length of the wire, the built-in tunnel barrier is estimated to be above the local gate *g1* (**Figure 3a-c**). To validate this the device was sliced open after the transport experiments using focused ion beam and inspected in TEM. **Figure 3b** shows an EDX map of a part of the device indicated by a red rectangle in **Figure 3a**. The built-in tunnel barrier, indicated in green, is indeed exactly above the local gate *g1*.

Due to the proximity of the built-in tunnel barrier, the local gate *g1* (which we from now on refer to as the *barrier gate*) is expected to show different gating effect on the device conductance compared to the local gates *g2* and *g3* (which are connected to act as a single local gate and referred to as the *normal gate*). **Figure 3d** shows the color plot of the two-point conductance G measured as a function of both the barrier gate and the normal gate voltage at 0 bias voltage (lock-in measurement) as well as schematic drawings illustrating the potential landscape in the device at a corresponding region in the plot.

Region (*i)* coincides with a high-conductance region: a high (more positive) gate voltage is applied to both the barrier gate and the normal gate, resulting in the bottom of the conduction band of both InSb and $Ga_{0.28}In_{0.72}Sb$ segment being pulled far below the Fermi level. Moving from (*i)* to (*ii)* in a straight line, the barrier-gate voltage remains unchanged while the normal-gate voltage decreases, pushing the conduction band of the InSb segment up towards the Fermi level. At (*ii)* the bottom of the conduction band of InSb is aligned with the Fermi level while the bottom of the barrier conduction band remains below the Fermi level. By decreasing the normal-gate voltage even further, we reach region (*iii)* in which the conduction band bottom of InSb is pushed above the Fermi level and the nanowire is not conducting.

If we start from region (*i)* and move towards (*iv)* in a straight line, the normal gate voltage remains high while the barrier gate voltage decreases, pushing the conduction band of the barrier segment up towards the Fermi level. At (*iv)* the bottom of the conduction band of the barrier is aligned with the Fermi level. If the barrier-gate voltage is reduced even further, conductance is zero (region (*v)*).

Note that the onset of transport in region (*iv)* happens at a significantly higher value of applied voltage (barrier-gate voltage ~2 V) than the onset of transport in region (*ii)* (normal-gate voltage ~0.2 V), indicating that the bottom of the conduction band of the nanowire segment just above the tunnel barrier is higher than the bottom of the conduction band of the nanowire segment above the normal gate, as expected if a built-in barrier is present in the nanowire section above the barrier gate. In region (*vi)*, all the local back gates are set to the value (~ 0.2 V) which aligns the bottom of the InSb conduction band to the Fermi level. Here, the height of the built-in tunnel barrier is expected to correspond to the actual conduction band offset between the $Ga_{0.28}In_{0.72}Sb$ and InSb nanowire segments.

The above analysis is based on a simple assumption: each local gate mainly tunes the potential in the nanowire segment exactly above it, and the cross-coupling between individual gates is negligible. The fact that the threshold voltage for barrier gate (normal gate) does not alter significantly while the normal gate (barrier gate) is tuned, supports this assumption.

To extract the width and the height of the built-in $Ga_{0.28}In_{0.72}Sb$ tunnel barrier, we measure current I as a function of bias voltage $V_{bias}$ and gate voltage $V_{gate}$ (gates *g1, g2* and *g3* connected, acting as a single gate). The blue (red) region in the resulting color plot (**Figure 4a**) corresponds to the region of low (high) current. The asymmetric behavior between positive and negative biasing is most likely a consequence of the asymmetric biasing effect of the circuit (possible explanation of the asymmetric I-V behavior can be found in **SI-9**).

A line cut taken at $V_{gate}$=1.4 V is plotted as a blue dotted line in **Figure 4b**. The red solid line in **Figure 4b** represents a theoretical fit to the experimental data (see **SI-7** for a detailed explanation of the WKB model employed) calculated assuming a square-shaped barrier potential. The fitted barrier width (19.2 ± 0.2 nm) is in excellent agreement with the value extracted from EDX (**Figure 3b**) of the same device. The fitted barrier height at $V_{gate}$=1.4 V equals 44.2 ± 0.3 meV. The black solid line in **Figure 4b** corresponds to a WKB model fit of the experimental I-V curve calculated assuming a Gaussian-shaped potential barrier. The width (height) of the Gaussian barrier at $V_{gate}$=1.4 V equals 21 ± 1 nm (45 ± 2 meV). Since both the square- and Gaussian-shaped barrier potential fit well with the experimental data, we conclude that the extracted values of barrier width and height do not depend on the details of the barrier potential profile. Hence, in the remainder of this paper, the WKB modelling is based on a square-shaped barrier potential. The inset of **Figure 4b** shows that a good fit to the experimental I-V curve (blue dots) can only be achieved if a barrier width of ~ 20 nm is assumed. The orange (purple) solid line corresponds to the best WKB fit of a 14 nm (30 nm) wide potential barrier, using only the barrier height as the free parameter. The corresponding barrier height is 63 ± 1 meV (30 ± 1 meV) for the 14nm (30 nm) barrier width.

**Figure 4c-d** shows the barrier width and height (extracted from WKB models of the I-V traces in **Figure 4a,** see **SI-8**) as a function of gate voltage $V_{gate}$. **Figure 4c** clearly demonstrates that, as the value of $V_{gate}$ increases from 0.6 V to 1.7 V, the effective barrier height decreases from 120 meV to 30 meV. (For $V_{gate}$<0.6V, the conduction band bottom approaches the Fermi level and the non-uniform gating effect becomes more pronounced, causing the breakdown of the simple square-shape model; see **SI-9**.) **Figure 4d** shows that, for the same range of $V_{gate}$, the barrier width remains roughly the same (~ 20nm), while the height is changed by a factor of 4. The fact that the barrier width does not depend on the value of the gate voltage applied indicates that we are indeed measuring a material-defined tunnel barrier; the width of an electrostatic barrier is expected to alter significantly under influence of $V_{gate}$. If we extrapolate the data shown in **Figure 4c**, we can roughly estimate the height of the built-in $Ga_{0.28}In_{0.72}Sb$ barrier at $V_{gate}$=0.2 V (which corresponds to the actual conduction band offset between $Ga_{0.28}In_{0.72}Sb$ and InSb segments; region (*vi)* in **Figure 3d**) to be ~200 meV (this value is in good agreement with the bulk values: the conduction band offset between the bulk InSb and GaSb at 0 K is 577 meV,[51,52] simple interpolation for *x*(Ga)=0.28 gives the conduction band offset ~ 160 meV).

A similar analysis of a nanowire device containing a 20 nm wide $Ga_{0.15}In_{0.85}Sb$ segment is shown in the **SI-10.** From the measured I-V characteristics and the corresponding WKB-model-fits, we determine the conduction band offset between $Ga_{0.15}In_{0.85}Sb$ and InSb segments to be ~75 meV (in good agreement with the bulk values; simple interpolation for *x*(Ga)=0.15 gives the conduction band offset ~ 85 meV).

In summary, we demonstrate the growth of defect-free InSb nanowires with composition- and size- tunable $Ga_xIn_{1-x}Sb$ barriers. The width and the height of the material-defined tunnel barriers is extracted from WKB-model-fits to experimental I-V traces and the conduction band offset between InSb and $Ga_xIn_{1-x}Sb$ (for *x*(Ga) = 0.15 and *x*(Ga)=0.28) is determined. Implementation of these InSb/$Ga_xIn_{1-x}Sb$/InSb nanowire heterostructures in the next generation of Majorana-detection devices can significantly improve the visibility of the topological ZBP.

**Supporting Information**.

Growth of the InSb/$Ga_xIn_{1-x}Sb$/InSb nanowires, extracting the length of the barriers from HRTEM images, transverse EDX line scan, strain quantification of a $Ga_{0.15}In_{0.85}Sb$ segment, statistics on the position of the barrier within the nanowire, detailed device fabrication recipe, explanation of the I-V fitting model, additional WKB fits of I-V curves, a possible explanation of the asymmetric I-V behavior, transport measurements of an InSb/$Ga_{0.15}In_{0.85}Sb$/InSb nanowire device (PDF).

**Funding Sources**

This work has been supported by the Netherlands Organization for Scientific Research (NWO), Foundation for Fundamental Research on Matter (FOM), European Union Seventh Framework Programme, European Research Council (ERC), Office of Naval Research (ONR) and Microsoft Corporation Station Q.

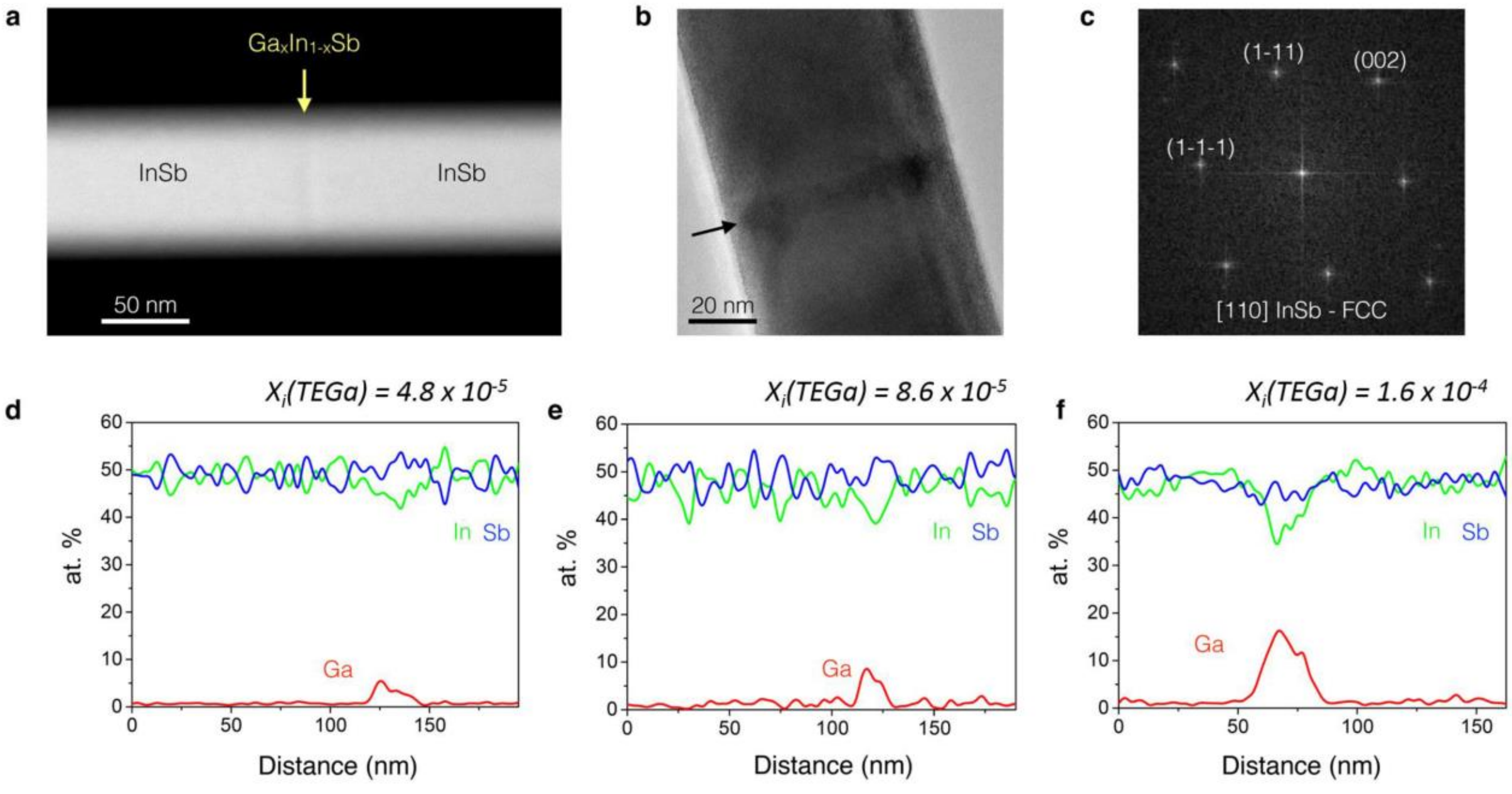

**Figure 1.** InSb nanowires with $Ga_xIn_{1-x}Sb$ axial segments. (a) HAADF-STEM of an InSb/$Ga_xIn_{1-x}Sb$/InSb nanowire heterostructure. The yellow arrow indicates the thin $Ga_xIn_{1-x}Sb$ segment. (b) A HRTEM image, taken in the [110] zone axis, of a part of an InSb nanowire containing the $Ga_xIn_{1-x}Sb$ segment, indicated by a black arrow. (c) A FFT of the HRTEM image shown in (b) reveals the pure zinc-blende crystal structure. (d)-(f) Tuning the chemical composition of the $Ga_xIn_{1-x}Sb$ segments. By increasing the TEGa molar fraction, *$X_i(TEGa)$*, while keeping the growth time fixed, we increase the Gallium content in the $Ga_xIn_{1-x}Sb$ segments while keeping the segment length fixed. From the EDX line scans we can extract the chemical composition and the length of the $Ga_xIn_{1-x}Sb$ segments: (d) $Ga_{0.1}In_{0.9}Sb$, 20 nm (e) $Ga_{0.15}In_{0.85}Sb$, 20 nm (f) $Ga_{0.28}In_{0.72}Sb$, 20 nm.

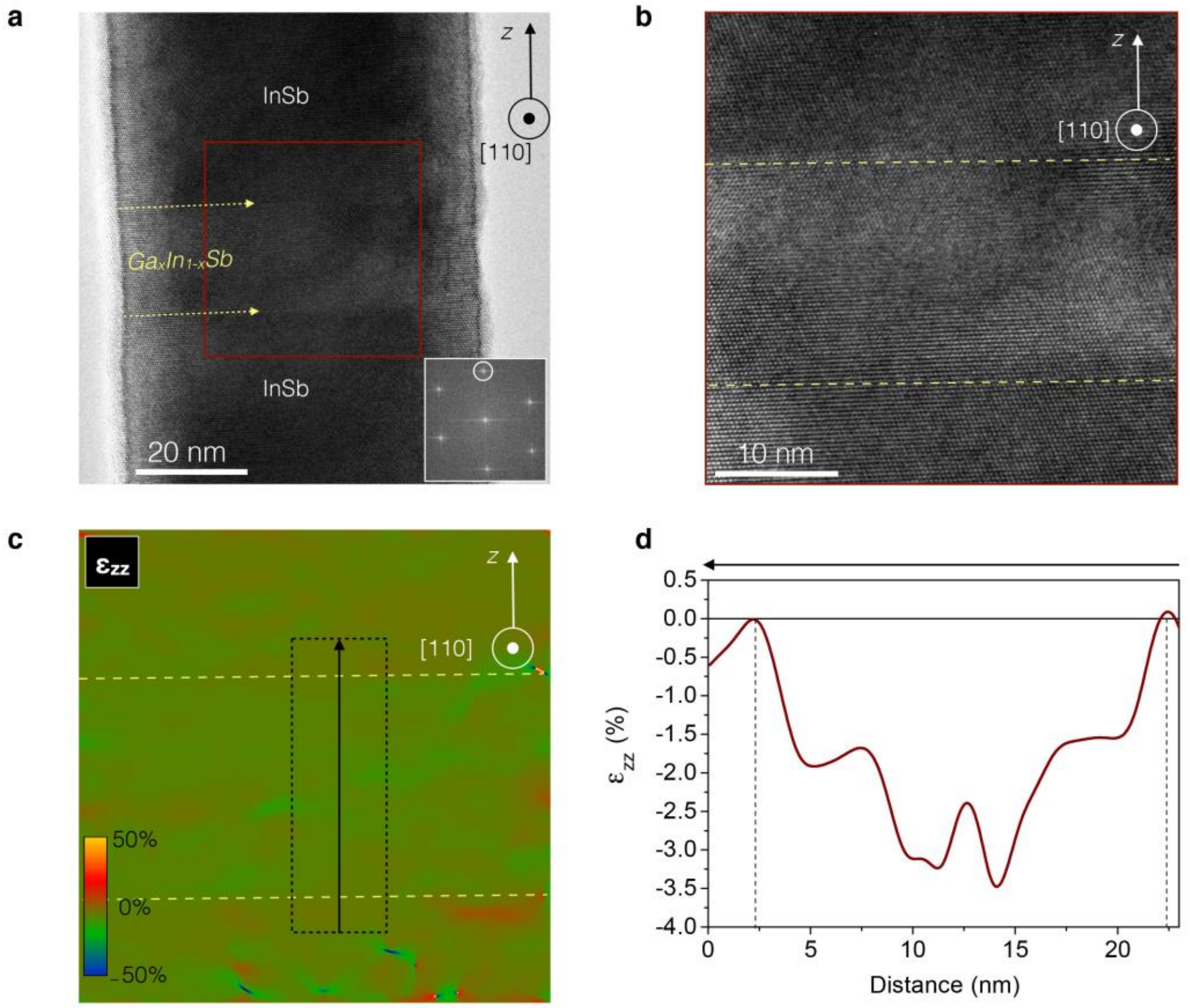

**Figure 2.** Strain quantification by Geometrical Phase Analysis (GPA). (a) A HRTEM image, taken in the [110] zone axis, of a nanowire containing a 20 nm thick $Ga_{0.28}In_{0.72}Sb$ barrier, outlined by yellow dashed arrows. Inset shows the corresponding FFT. The encircled spot corresponds to the (1-11) set of planes. The nanowire growth direction *z* is [1-11], as indicated by a black arrow. (b) A zoom-in on a region indicated by a red square in (a). (c) $\varepsilon_{zz}$ component of the strain tensor as calculated from the GPA applied to the (1-11) planes of the HRTEM image in (b). The $Ga_{0.28}In_{0.72}Sb$ segment is compressively strained along the *z* direction with respect to the InSb reference region. (d) The strain profile integrated along the direction indicated by the black arrow in (c). The average value of compressive strain in the *z*-direction is around -2%.

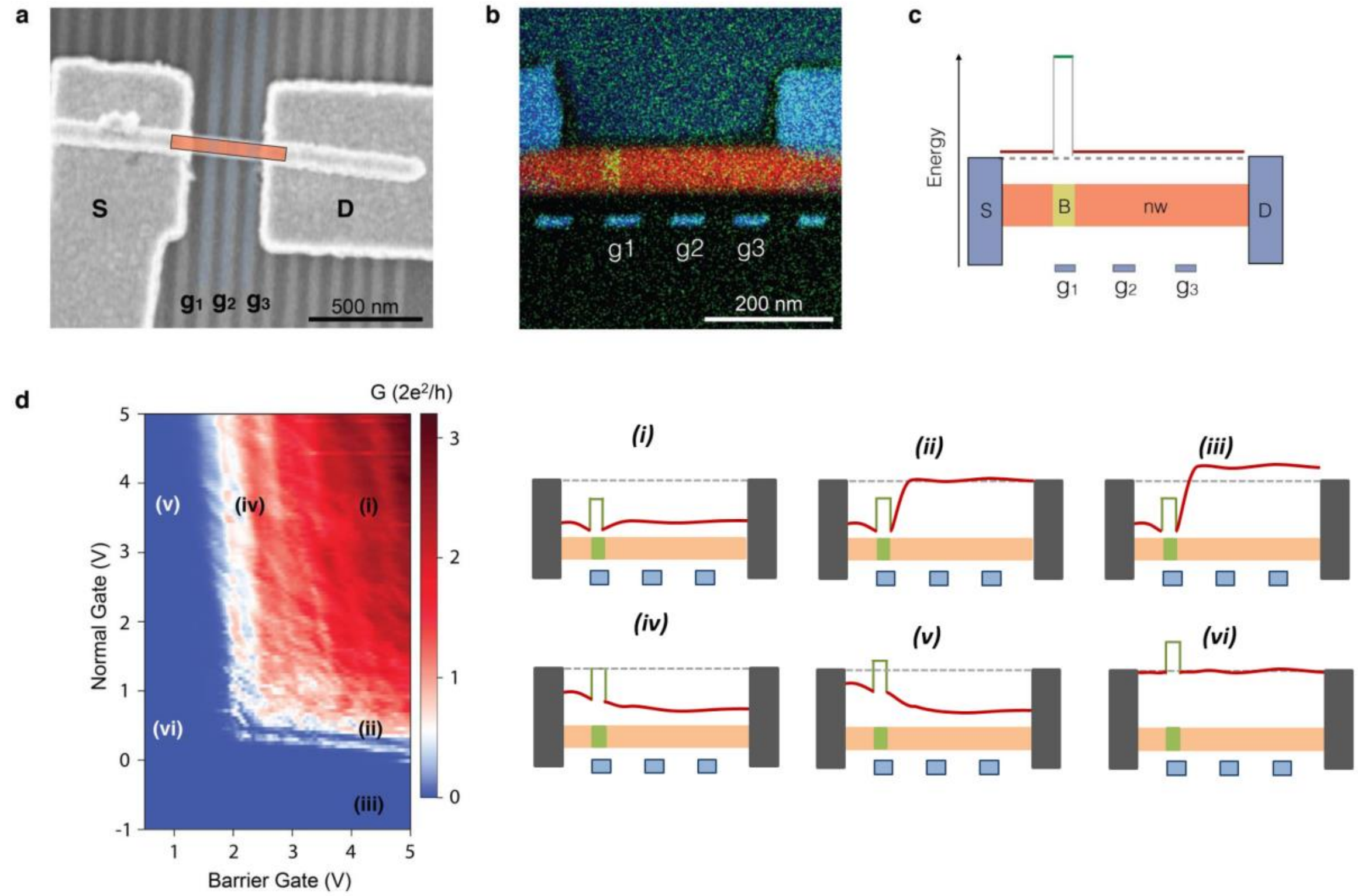

**Figure 3.** Transport measurements of an InSb/$Ga_{0.28}In_{0.72}Sb$/InSb nanowire device. (a) A top-view SEM image of the device. Part of the device (false colored in red) was sliced open in focused ion beam and inspected sideways in TEM/EDX. (b) An EDX map of the region indicated by a red rectangle in (a). The EDX analysis confirms that the position of the built-in barrier (indicated in green) is right above the local back-gate *g1*. (c) A simplified schematic drawing of the device and its potential landscape. (d) Color plot of conductance *G* as a function of barrier gate (*g1*) voltage and normal gate (*g2* and *g3* connected, acting as a single gate) voltage. Schematic drawings on the right illustrate potential landscape (*i.e.* bottom of the conduction band of the InSb and $Ga_{0.28}In_{0.72}Sb$ segments) at different regions in the color plot. The onset of transport in region (*iv)* happens at a significantly higher value of applied voltage (barrier gate voltage ~2 V) than the onset of transport in region (*ii)* (normal gate voltage ~0.2 V), indicating that a built-in barrier is present in the nanowire, in the region above the barrier-gate (*g1*).

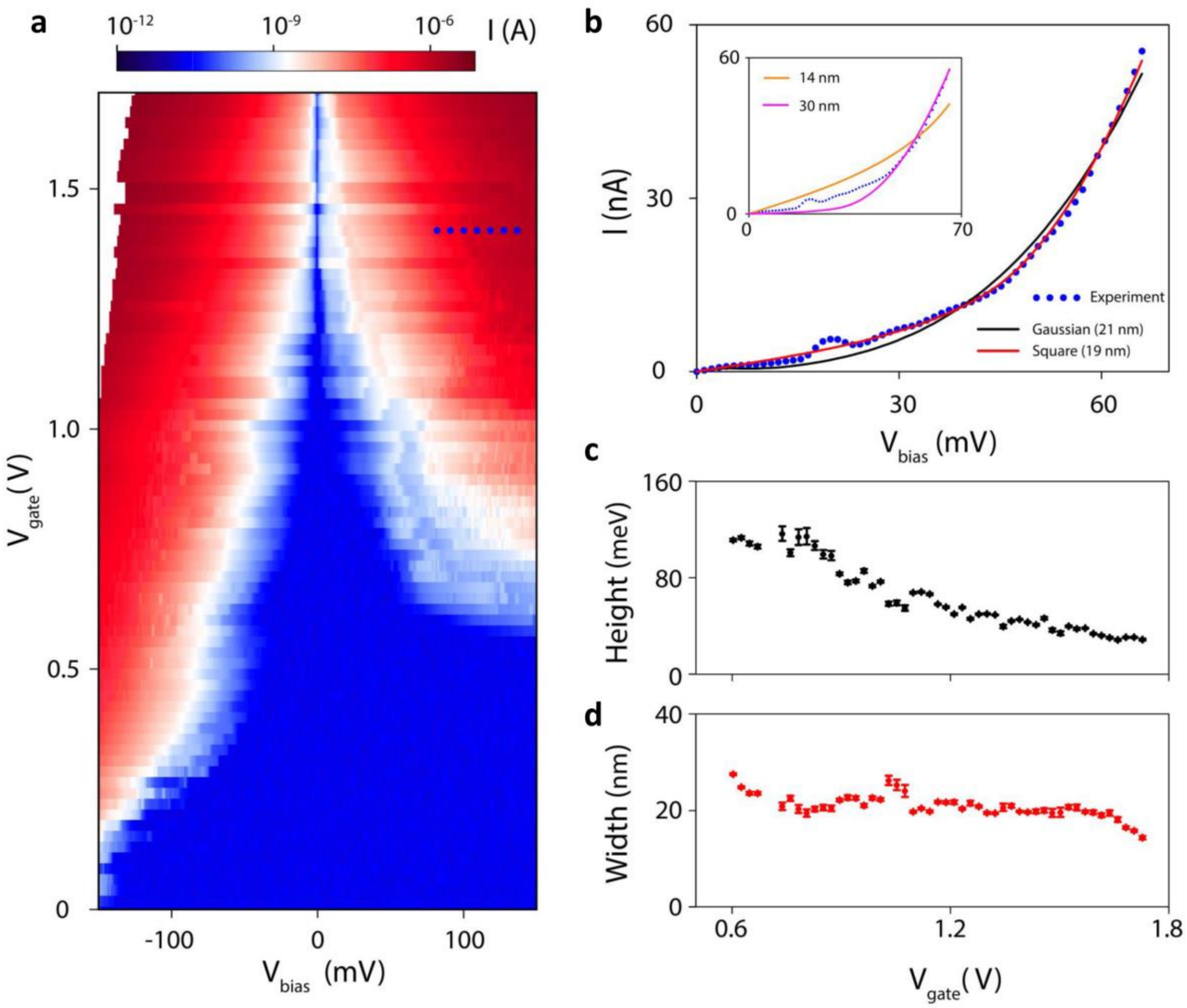

**Figure 4.** Extracting the built-in barrier height and width. (a) Color plot of current I as a function of bias voltage $V_{bias}$ and gate voltage $V_{gate}$ (local gates *g1, g2* and *g3* connected, acting as a single gate). (b) Experimental I-V trace (blue dots) taken at $V_{gate}$=1.4 V (indicated by a blue dotted line in (a)) and WKB fits calculated assuming a square-shaped (red solid line) and Gaussian-shaped (black solid line) barrier potential. The fitted barrier width (height) is 19.2 ± 0.2 nm (44.2 ± 0.3 meV) for the square-shaped barrier. In case of the Gaussian-shaped barrier, the fitted barrier width (height) is 21 ± 1 nm (45 ± 2 meV). The inset shows the best WKB fits obtained assuming the barrier width of 14 nm (orange solid line) and 30 nm (purple solid line). The good fit to the experimental data can only be obtained assuming the ~20 nm barrier width. (c) The barrier height as a function of gate voltage $V_{gate}$. The effective barrier

height decreases from 120 meV to 30 meV as the value of $V_{gate}$ increases from 0.6 V to 1.7 V. (d) The barrier width as a function of gate voltage $V_{gate}$. For the same $V_{gate}$ range as in (c), the barrier width remains roughly the same (~20 nm), while the height changes by a factor of 4. In both (c) and (d) the square-shaped barrier potential is used.

# Supporting Information

# InSb Nanowires with Built-In $Ga_xIn_{1-x}Sb$ Tunnel Barriers for Majorana Devices

*Diana Car,[1,ǂ] Sonia Conesa-Boj,[2,ǂ] Hao Zhang,[2,ǂ] Roy L. M. Op het Veld,[2] Michiel W. A. de Moor,[2] Elham M. T. Fadaly,[2] Önder Gül,[2] Sebastian Kölling,[1] Sebastien R. Plissard,[3] Vigdis Toresen,[2] Michael T. Wimmer,[2] Kenji Watanabe,[4] Takashi Taniguchi,[4] Leo P. Kouwenhoven,[2] Erik P. A. M. Bakkers[*,1,2]*

[1] Department of Applied Physics, Eindhoven University of Technology, P.O. Box 513, 5600 MB Eindhoven, the Netherlands

[2]Kavli Institute of Nanoscience, Delft University of Technology, 2628CJ Delft, the Netherlands

[3] CNRS-LAAS, 7 avenue du Colonel Roche, F-31400 Toulouse, France

[4] Advanced Materials Laboratory, National Institute for Materials Science, 1-1 Namiki, Tsukuba 305-0044, Japan

[ǂ]These authors contributed equally.

*Corresponding author: e.p.a.m.bakkers@tue.nl

# Section 1 (S1): Growth of InSb/$Ga_xIn_{1-x}Sb$/InSb nanowires

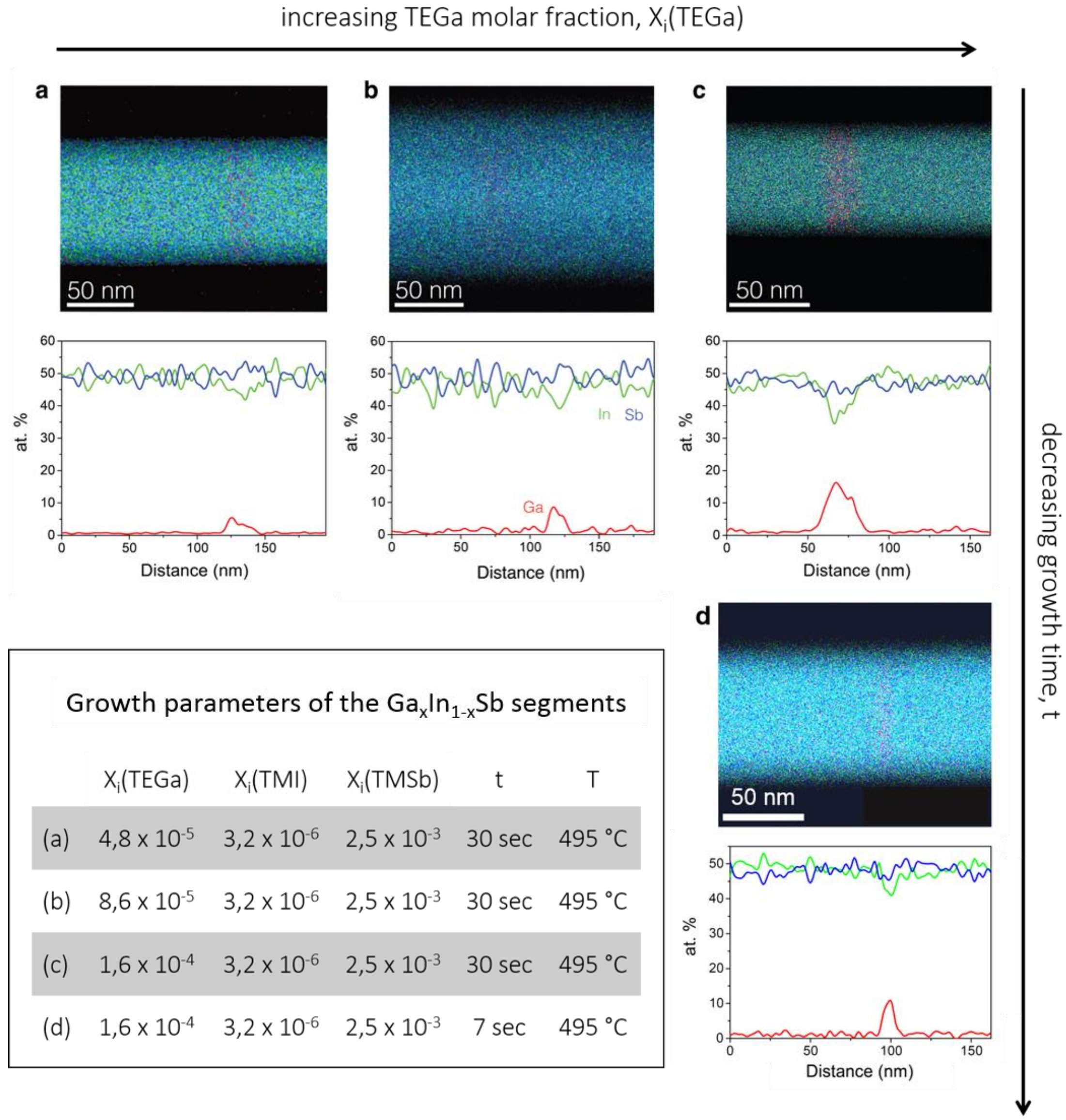

Growth parameters of the $Ga_xIn_{1-x}Sb$ segments

| | $X_i$(TEGa) | $X_i$(TMI) | $X_i$(TMSb) | t | T |
|---|---|---|---|---|---|
| (a) | 4,8 x $10^{-5}$ | 3,2 x $10^{-6}$ | 2,5 x $10^{-3}$ | 30 sec | 495 °C |
| (b) | 8,6 x $10^{-5}$ | 3,2 x $10^{-6}$ | 2,5 x $10^{-3}$ | 30 sec | 495 °C |
| (c) | 1,6 x $10^{-4}$ | 3,2 x $10^{-6}$ | 2,5 x $10^{-3}$ | 30 sec | 495 °C |
| (d) | 1,6 x $10^{-4}$ | 3,2 x $10^{-6}$ | 2,5 x $10^{-3}$ | 7 sec | 495 °C |

Figure S4. Tuning the composition and length of $Ga_xIn_{1-x}Sb$ segments. From the EDX maps and line scans shown in (a)-(d) we can extract the composition and length of respective $Ga_xIn_{1-x}Sb$ segments. (a)-(c) By increasing the TEGa source flow while keeping the growth time fixed, we increase the Gallium content in the $Ga_xIn_{1-x}Sb$ segments while keeping the segment length fixed: (a) $Ga_{0.1}In_{0.9}Sb$, 20 nm (b) $Ga_{0.15}In_{0.85}Sb$, 20 nm (c) $Ga_{0.28}In_{0.72}Sb$, 20 nm. By reducing the growth time from (c) 30 sec to (d) 7 sec, we reduce the length of the barrier from 20 nm to 7 nm. The chemical composition of the segment shown in (d) is $Ga_{0.21}In_{0.79}Sb$. The table in the inset shows the growth parameters used to synthesize the $Ga_xIn_{1-x}Sb$ segments of different samples.

InSb/$Ga_xIn_{1-x}Sb$/InSb nanowire heterostructures have been synthesized by Au-catalyzed Vapor-Liquid-Solid (VLS) growth mechanism in an Aixtron Metal Organic Vapor Phase Epitaxy (MOVPE) machine. The growth was catalyzed by 30 nm Au colloids dispersed on an (001) InP substrate. InP nanowires are used as stems for InSb nanowire growth (see D.Car et al. *Adv. Mater.* **2014**, *26* (28), 4875–4879). For all samples, the InSb bottom and top nanowire segments were grown at 495 °C using tri-methyl-indium (TMI) and tri-methyl-antimony (TMSb) with precursor molar fractions $X_i$(TMI)= 1.1 x $10^{-5}$ and $X_i$(TMSb)= 1.7 x $10^{-3}$, for 8 (6) minutes for the bottom (top) segment. The growth parameters used for the growth of the intermediate $Ga_xIn_{1-x}Sb$ segments of different samples are listed in the inset of **FigureS1**. In all the runs, the growth was interrupted for 2 min before and after the growth of the $Ga_xIn_{1-x}Sb$ segments to adjust the flow in the gas lines and ensure the abruptness of the interfaces.

## Section 2 (S2): Extracting the length of the barriers from HRTEM images

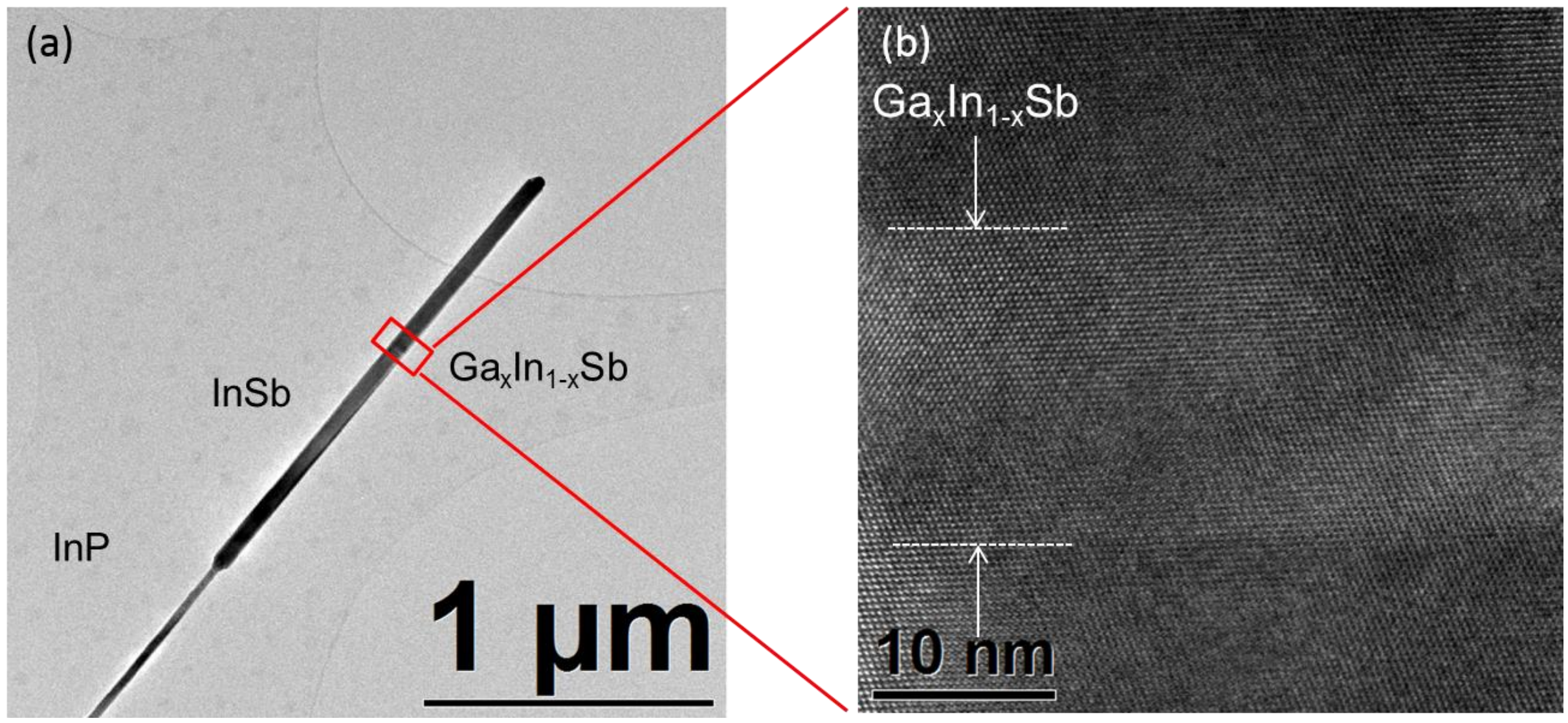

Figure S5. The length of the barrier as extracted from HRTEM image. (a) Bright-field, low-magnification TEM image of a nanowire from sample grown with the highest Ga molar fraction for 30 sec ($Ga_{0.28}In_{0.72}Sb$, 20 nm; Figure S1c). The nanowire is uniform in diameter and non-tapered. (b) A HRTEM image of a nanowire segment containing the barrier. The $Ga_xIn_{1-x}Sb$ segment appears brighter in contrast in HRTEM images. Measured barrier length is 18 nm, which is in good agreement with the value obtained from EDX line scan (~20 nm). In addition, from HRTEM image we can see that there are no misfit dislocations present.

# Section 3 (S3): Transverse EDX line scan shows no radial overgrowth

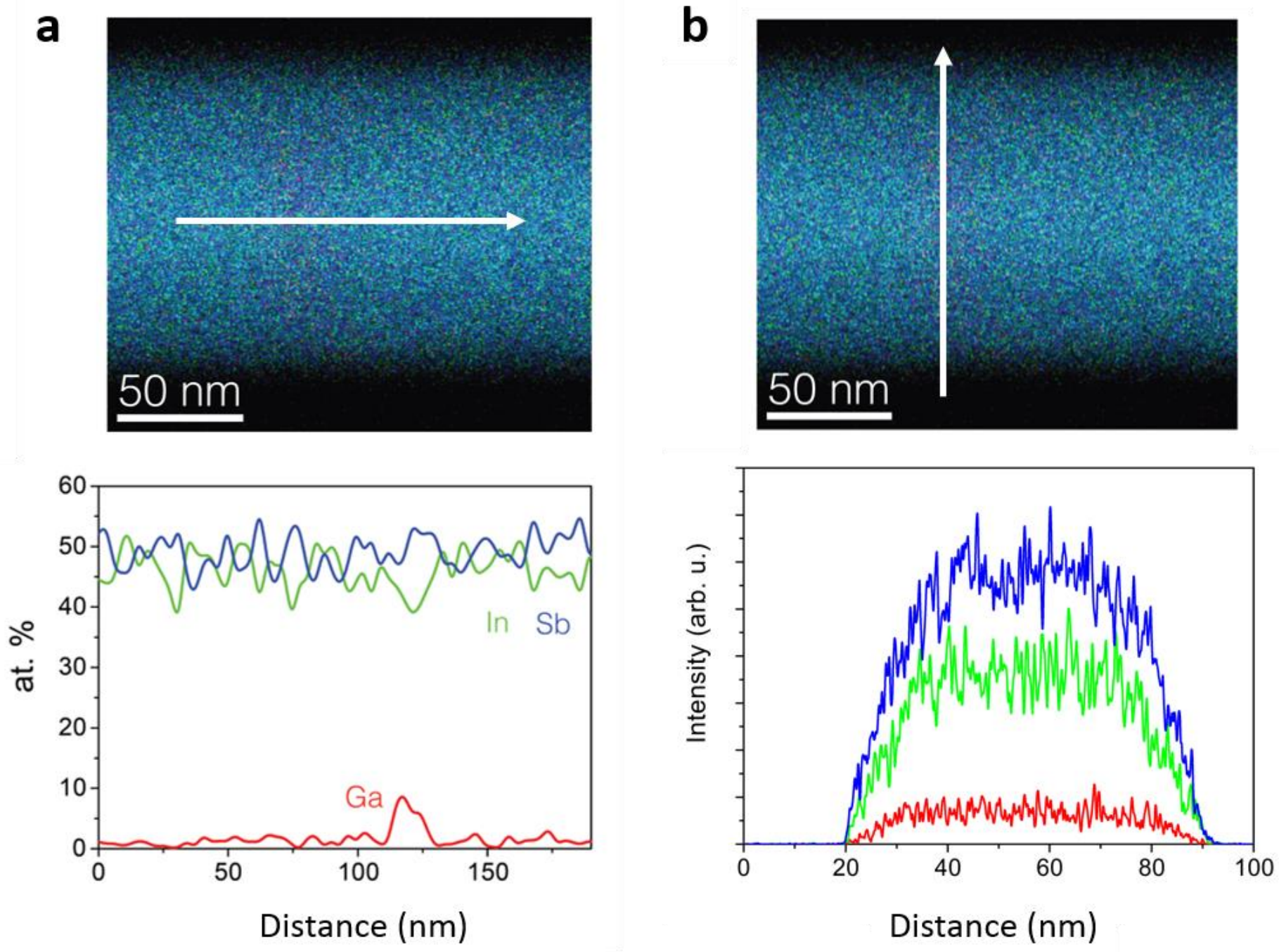

Figure S6. (a) Lower panel: EDX line scan taken along the axial direction (indicated by a white arrow in the XEDS map in the upper panel), revealing the barrier composition $Ga_{0.15}In_{0.85}Sb$ and thickness 20 nm. (b) Lower panel: EDX line scan taken along the radial direction (indicated by a white arrow in the EDX map in the upper panel) of the same $Ga_{0.15}In_{0.85}Sb$ segment. The barrier spans the full diameter of the nanowire.

## Section 4 (S4): Strain quantification of a $Ga_{0.15}In_{0.85}Sb$ segment

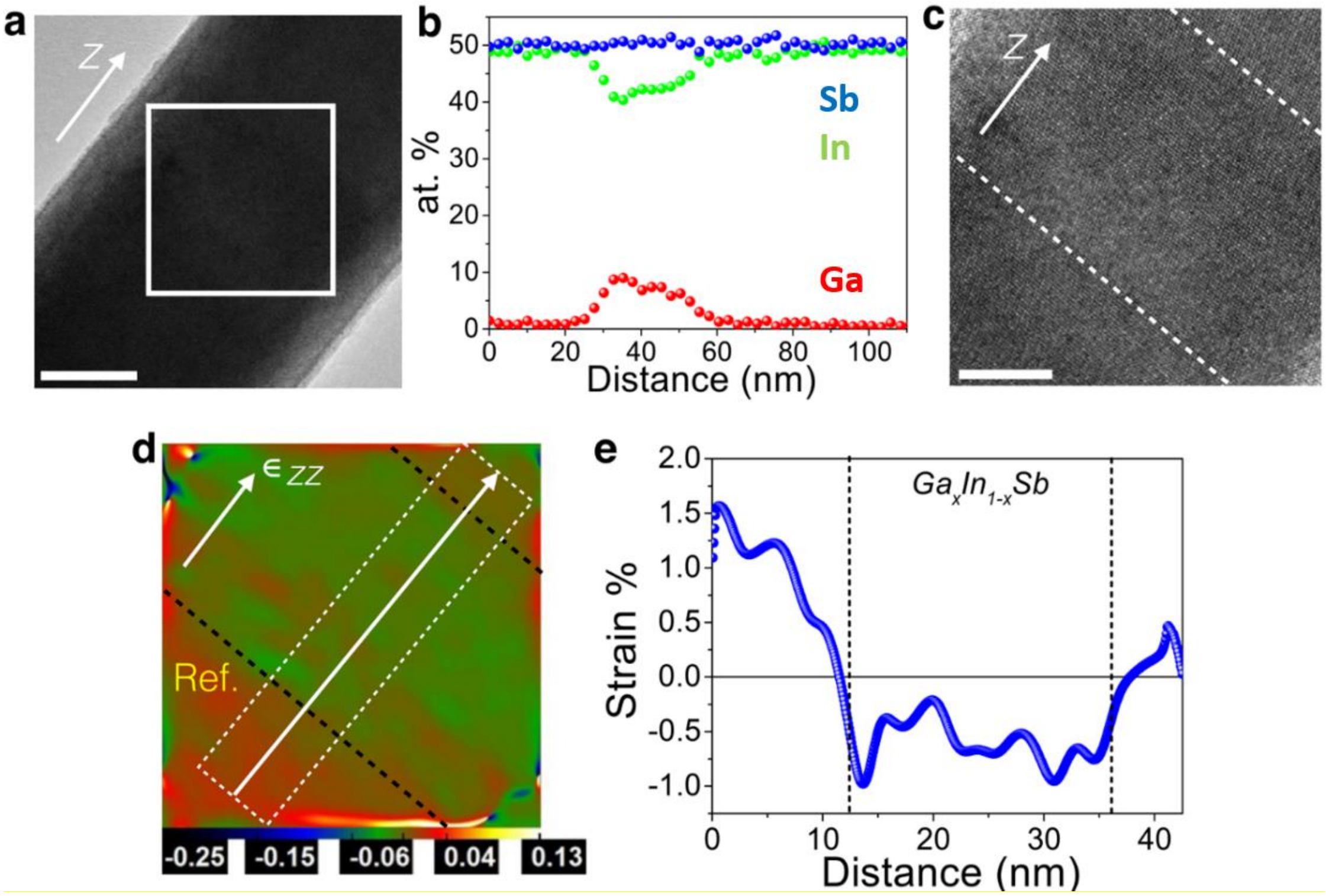

Figure S4. Strain in the InSb-$Ga_{0.15}In_{0.85}Sb$-InSb axial nanowire heterostructure. (a) Bright field TEM image of the nanowire segment containing the barrier. The arrow indicates the nanowire growth direction. Scale bar corresponds to 20 nm. (b) EDX line scan taken along the nanowire growth direction shows the presence of a 20 nm thick $Ga_{0.15}In_{0.85}Sb$ barrier. (c) High resolution TEM image of the region indicated by a square in (a). Scale bar corresponds to 10 nm. (d) Geometrical phase analysis applied to the (1-11) planes of the HRTEM image shown in (c).The $Ga_{0.15}In_{0.85}Sb$ segment is compressively strained with respect to the InSb reference (Ref.) region. (e) The strain profile integrated along the direction indicated by a white arrow in (d). The average value of strain in the barrier is around ~-0.8%.

## Section 5 (S5): Statistics on the position of the barrier within the nanowire

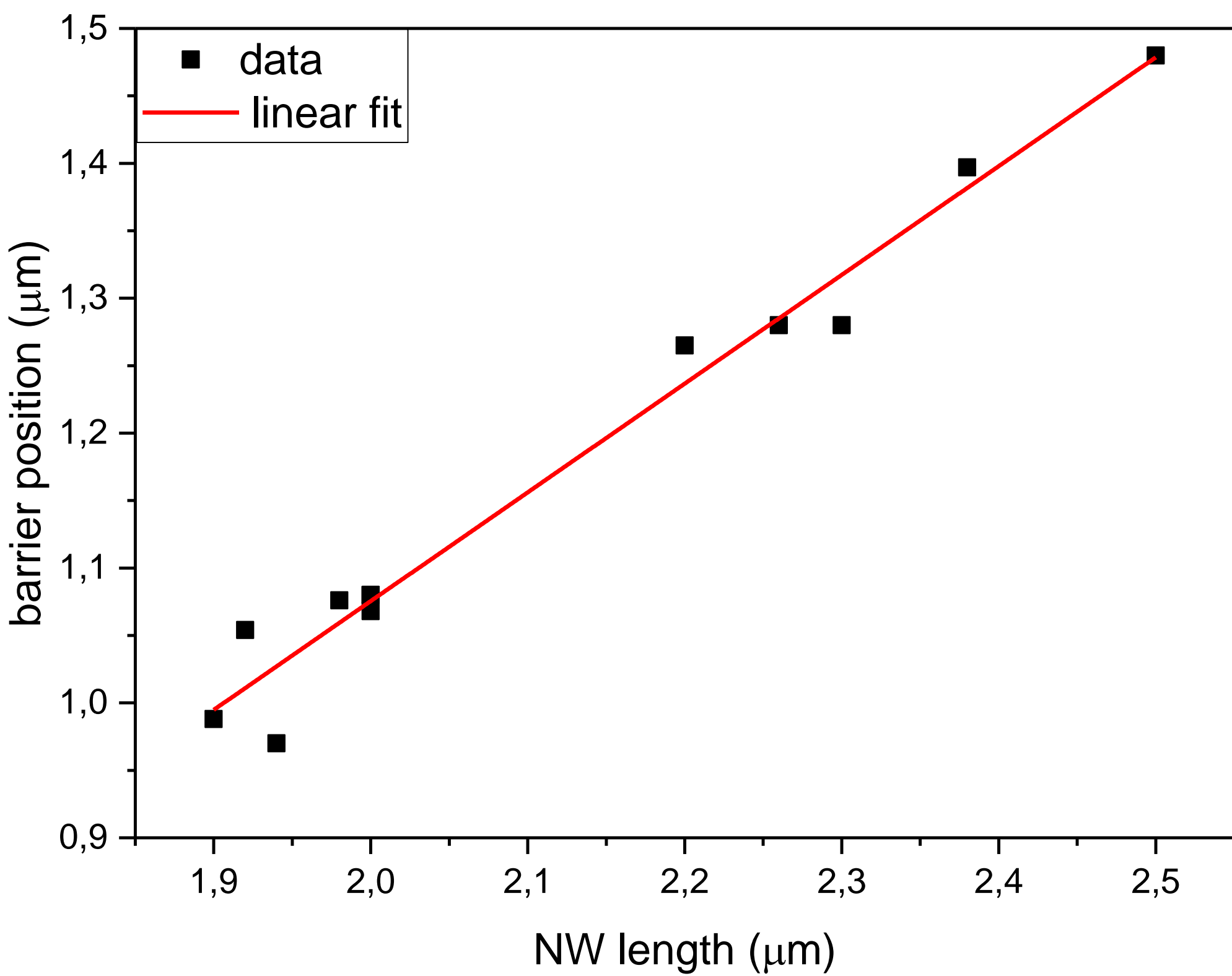

Figure S5. Statistical analysis of the sample containing the highest amount of Gallium ($Ga_{0.28}In_{0.72}Sb$) incorporated in a 20 nm thick barrier. The analysis was done to facilitate device fabrication by determining the position of the built-in barrier within the nanowire. The position of the barriers depends on the total length of the InSb wire. The spread in InSb NW growth rate is probably induced by fluctuations in Au catalyst density. The plot of barrier position (distance from the bottom of the InSb nanowire to the barrier) vs InSb nanowire length follows a linear law (*y=a+bx; a=(-0.54±0.09)μm; b=0.81±0.04*; the coefficient of determination (R-square value) is 0.97) and enables accurate prediction of the barrier position. A total of 11 nanowires was analyzed.

# Section 6 (S6): Detailed device fabrication recipe

**Substrate Cleaning**

Acetone (10 minutes), IPA (10 minutes), Oxygen plasma (10 minutes, pressure: 1mbar, power 600 watts).

**Fabrication of Local Gates**

Spin PMMA 950KA2 at 3000 rpm, bake at 175 °C for 15 minutes

Write local gate patterns (50 nm wide gates, 50 nm spacing) using e-beam lithography (dose: 1400 $\mu C/Cm^2$)

Developing for 60 seconds in MIBK (4-Methyl-2-pentanone): IPA, ratio 1:3

Rinse in IPA for 60 sec, blow dry

Evaporation of 5 nm Ti and 10nm Au

Lift-off in acetone overnight (with ultrasound for the first hour).

**Mechanical Exfoliation of Hexagonal Boron Nitride Flakes, Deterministic Transfer of the Flakes onto the Fine Bottom Gates**

**Nanowire Deposition**

Nanowires are transferred from the growth chip deterministically using a micromanipulator in SEM at 3 kV to the $Si/SiO_2$ chip between the alignment markers.

**Contact Deposition**

Spin PMMA 950(K) – A4 at 4000 rpm, bake at 175 C for 15 min

Writing contact patterns using e-beam lithography (dose: 1500 $\mu C/Cm^2$)

Developing for 60 sec in MIBK (4-Methylpentanon-2-one): IPA, ratio 1:3

Descum PMMA residues with oxygen plasma (60 sec, 1 mBar, 600 W)

Sulfur passivation:[1] Diluted ammonium polysulfide $(NH_4)_2S_x$ solution (3 ml of $(NH_4)_2S$ mixed with 290 mg sulfur powder then diluted with DI-water at a ratio of (1 : 200) for 30 min at 60 °C.

He Ion etching with a Kauffman ion source for 30 sec at 1.5-1.6 x $10^{-2}$ mbar

Evaporation of 10 nm Cr and 100nm Au

Lift-off in acetone overnight at room temperature.

## Section 7 (S7): Detailed explanation of the I-V fitting model

The height and width of the barrier are extracted from a least-square non-linear fit of the experimental I-V traces. The least-square non-linear fit is based on a simple theoretical model: When a voltage bias is applied, the bias drops linearly across the barrier width, resulting in a tilted barrier (**Figure S7c**). Then the tunneling transmission probability as a function of energy, T(E), is calculated using WKB approximation:

$$T(E, V_{bias}) = \exp(-\int_{x_L}^{x_R} 2\sqrt{2m(V(x) - E)/\hbar^2}\, dx)$$

where $x_L$ and $x_R$ are the classical turning points, $V$ is the barrier potential energy and $\hbar$ is the reduced Planck constant.

As illustrated in **Figure S7** two different barrier potential profiles are used: rectangular and Gaussian.

Rectangular barrier:

$$V(x) = V_{barr}\left[H\left(x + \frac{L}{2}\right) - H(x - \frac{L}{2})\right] + eV_{bias}\left(\frac{1}{2} - \frac{x}{L}\right)$$

Gaussian barrier:

$$V(x) = V_{barr}exp^{-\frac{x^2}{(2\sigma)^2}} + eV_{bias}\left(\frac{1}{2} - \frac{x}{L}\right)$$

Where H(x) is the Heaviside step function, $x_L$= min (x|V(x)=$E_F$), $x_R$= max(x|V(x)=$E_F$)

The width of the rectangular barrier is defined with respect to the shape at zero bias (**Figure S7a**) and is equal to the separation of the classical turning points $(x_L - x_R)$ for an electron at the Fermi level and the barrier height equals the difference between the top of the barrier and the Fermi level, $V - E_F$. The width of the Gaussian barrier (**Figure S7b**) is defined as 2 standard deviations ($2\sigma$) of a Gaussian at zero bias and the height as the difference between the peak height of the Gaussian and the Fermi level, $V_{barr} - E_F$

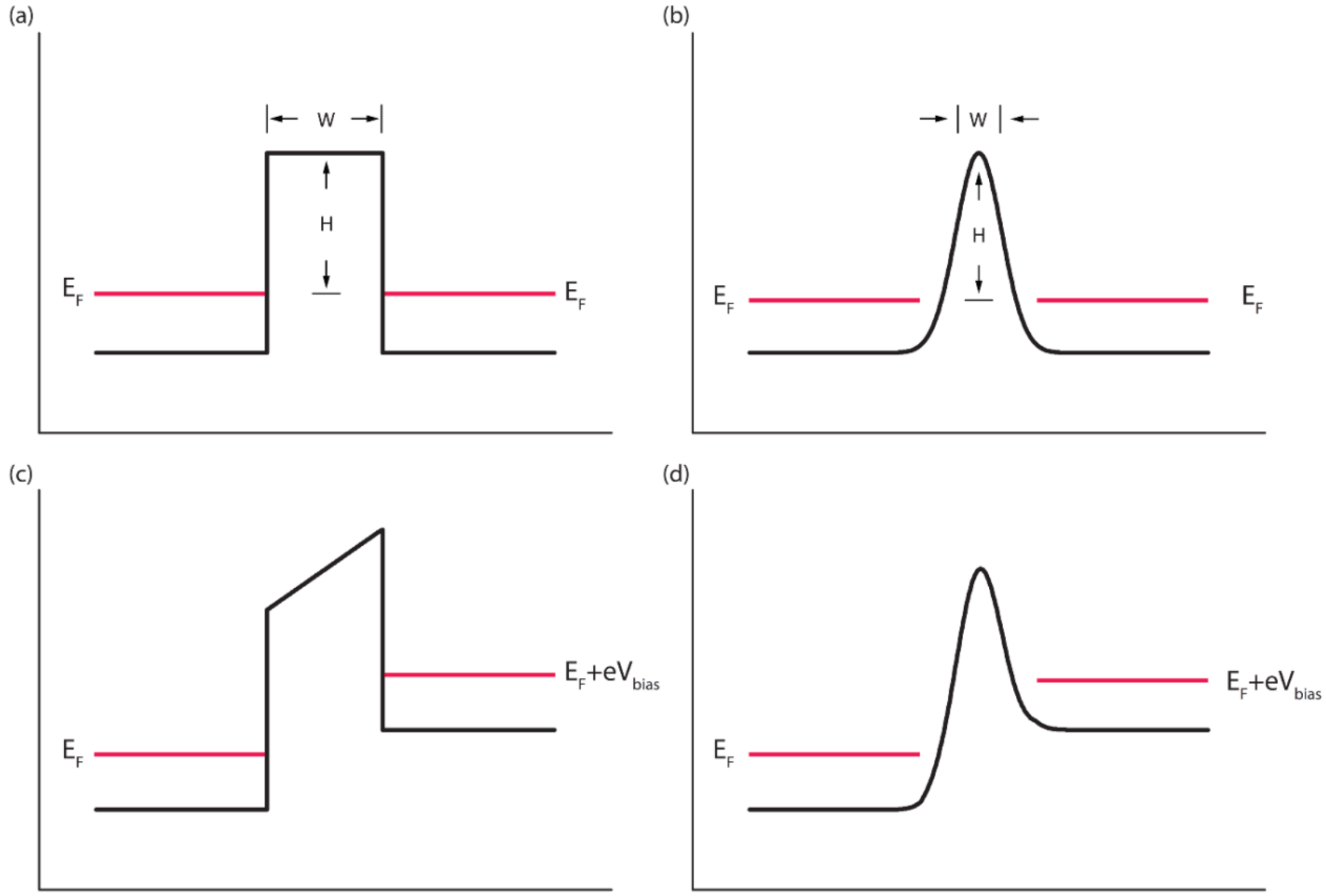

Figure S7. Schematics showing the potential landscape of the barrier region for the rectangular and Gaussian barriers assumed in the calculation of the transmission for the WKB fittings. Panel (a) and (b) show the barrier at zero bias voltage with the Fermi level indicated by red lines. Panel (c) and (d) show how the potential profile is modified at finite bias with a linear voltage drop across the barrier width.

After the transmission probability T(E) is calculated, the current I is obtained from the Landauer–Buttiker formula at zero temperature.

$$I(V_{bias}) = \frac{2e^2}{h} \int_{E_F}^{E_F+eV_{bias}} T(E, V_{bias}) dE$$

The assumption of zero temperature is valid due to the fact that the barrier height energy scale (>20meV) is much larger than the thermal energy scale (<1meV at 2 Kelvin).

## Section 8 (S8): Additional WKB fits of I-V traces

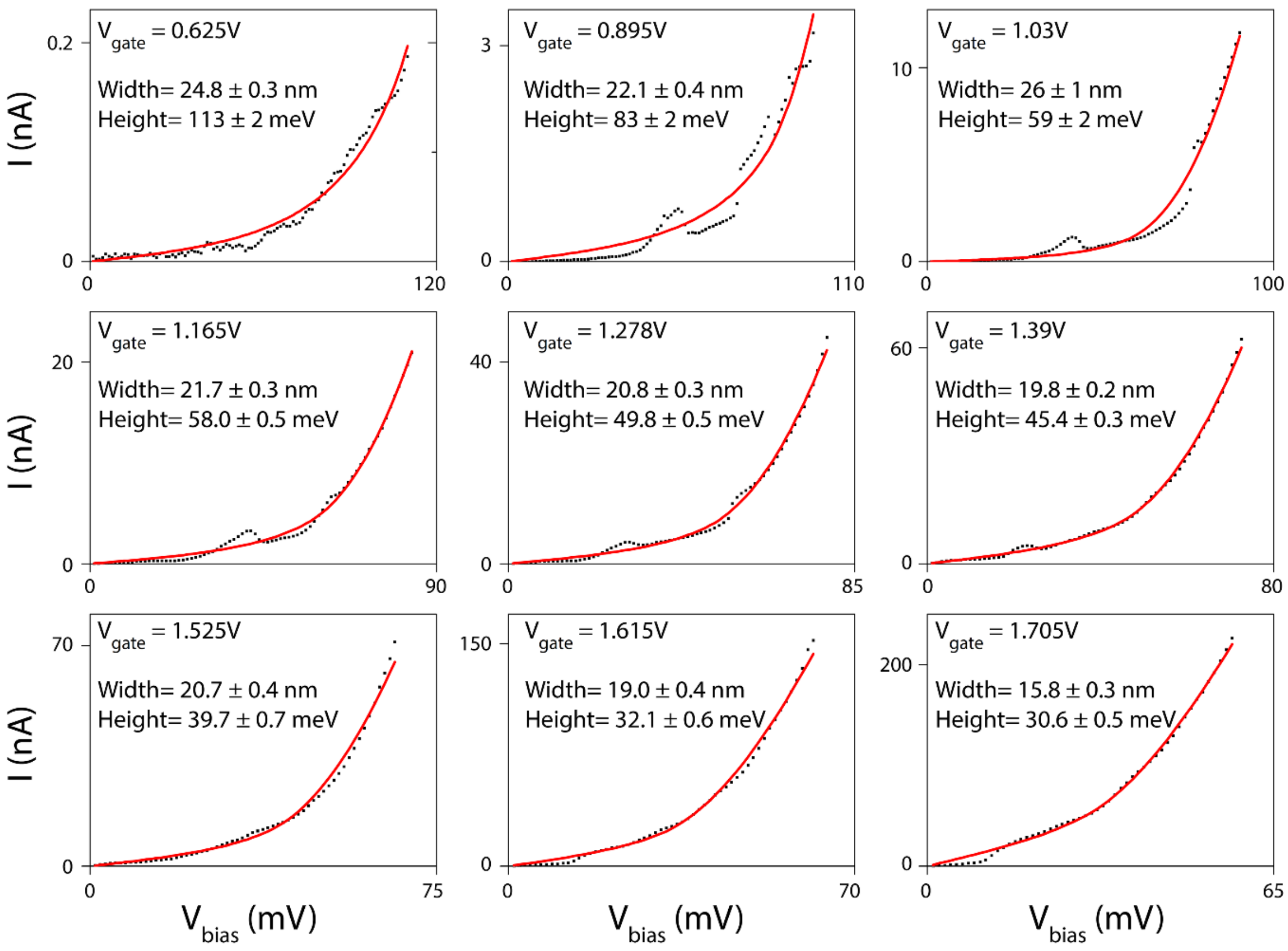

Figure S8. Representative WKB fitting curves, for I-V traces taken at gate-voltage values from 0.6 V to 1.7 V. As can be seen, the WKB model based on a simple, square-shaped potential profile shows excellent fit with experimental data for gate voltage values in the range from 0.625 V to 1.705 V. As the gate voltage increases from 0.625 V to 1.705 V, the current changes from 0.2 nA to 200 nA, *i.e.* by 3 orders of magnitude.

## Section 9 (S9): A possible explanation of the asymmetric I-V characteristics

The asymmetry of the I-V traces between positive bias and negative bias is more pronounced at low gate voltage values. We draw a possible explanation for this observation based on device asymmetry when the Fermi level lies below the conduction band. **Figure S9a** shows the potential landscape for zero bias voltage. **Figure S9b** depicts the case for negative bias. The bias voltage drops over the built-in barrier as well as the resistive nanowire section for which the Fermi level is below the conduction band. **Figure S9c** depicts the case for positive bias. There is an additional barrier apart from the built-in one which electrons have to tunnel through, causing an asymmetry for positive bias.

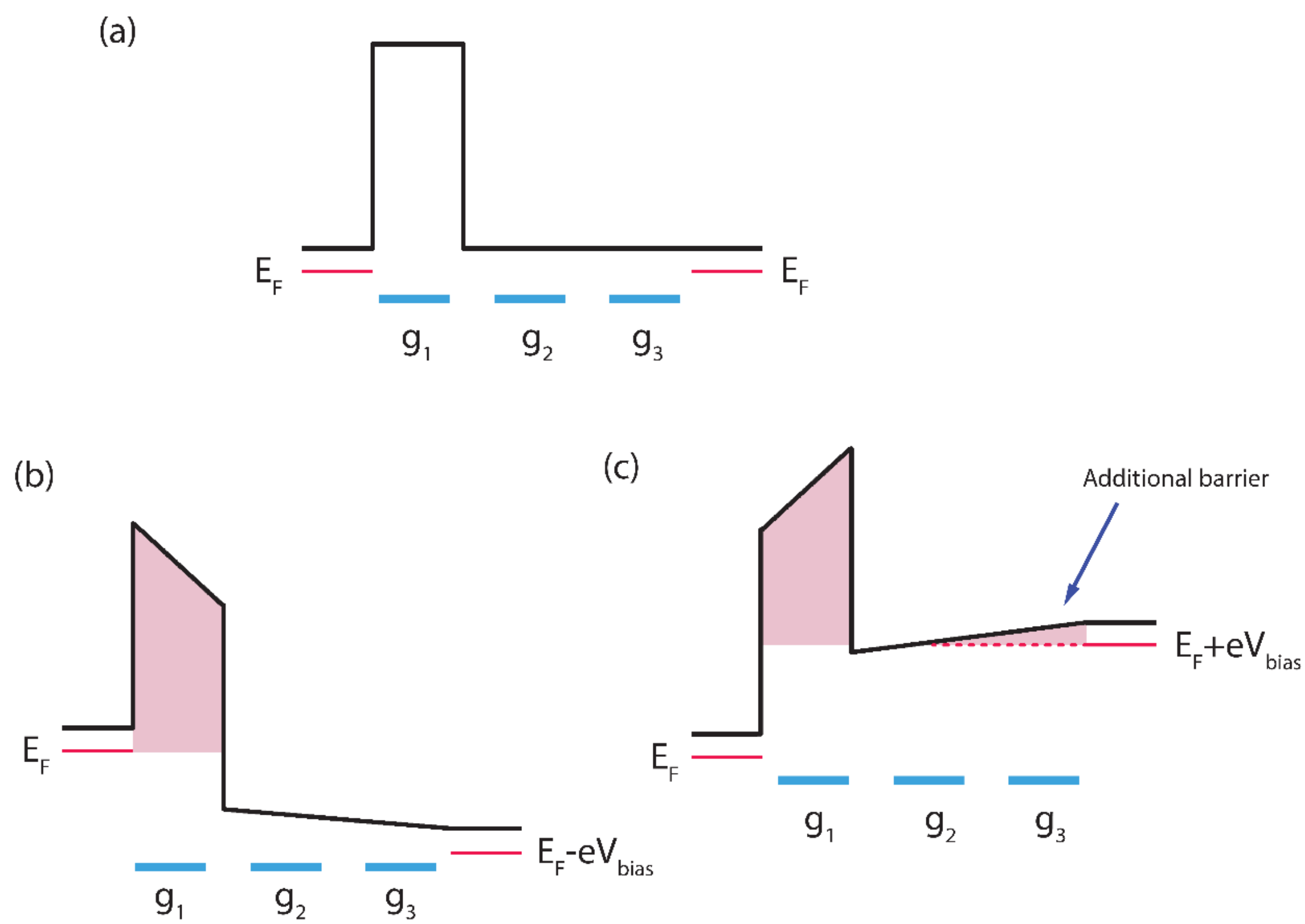

Figure S9. Potential landscape for less positive gate voltages. (a), (b) and (c) are for 0, negative and positive bias voltages, respectively. The three local back-gates are denoted as g1, g2 and g3.

We point out that this is a simplified model: In the vicinity of the conduction band bottom, the potential fluctuations due the local gates are more pronounced (e.g. the nanowire part in-between g1 and g2 is less capacitively coupled to the gates than the nanowire part right above the g1 or g2). Thus it is expected that our simple square-shaped barrier model no longer holds in this regime.

These effects (asymmetric biasing and potential fluctuations) can be minimized by applying more positive gate voltage to drag the conduction band bottom way below the Fermi level. In this case, the transport (which mainly happens near the Fermi level) is not sensitive to the details of the conduction band bottom, and our theory model fits to the data very well (**Figure 4** in the maintext).

## Section 10 (S10): Transport measurements of an InSb/$Ga_{0.15}In_{0.85}Sb$/InSb nanowire device.

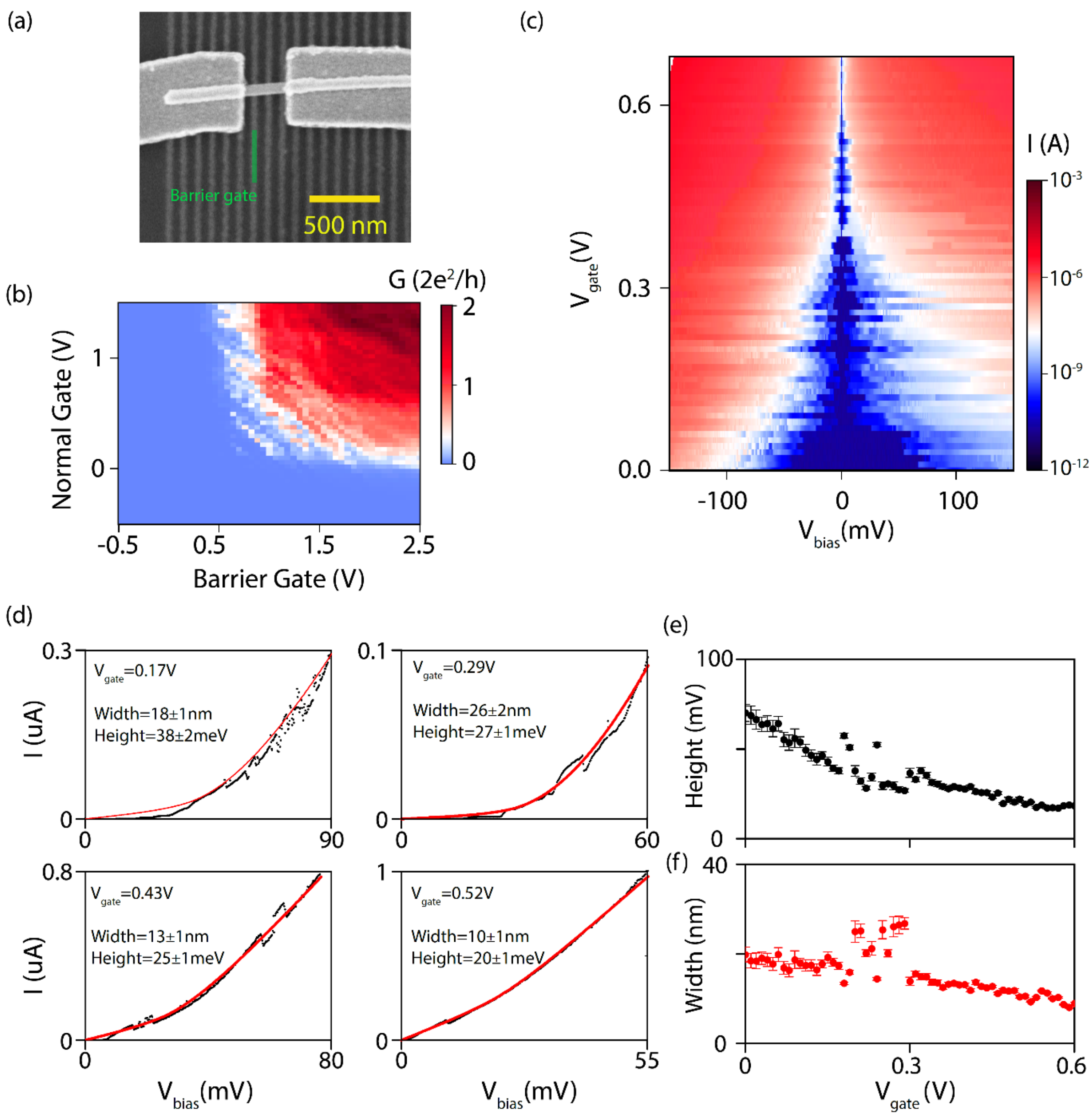

Figure S10: Transport measurements of an InSb/$Ga_{0.15}In_{0.85}Sb$/InSb nanowire device. The $Ga_{0.15}In_{0.85}Sb$ segment is expected to be 20 nm wide, as determined by TEM/EDX (Figure 1e in the main text). (a) Top-view SEM image of the device. The barrier gate is indicated by a green line. The normal gate is the local gate right next to the barrier gate (on the right-hand side). (b) Color plot of conductance G as a function of barrier gate and normal gate voltage, at 0 bias voltage (lock-in measurement). The threshold voltage difference between the barrier gate and the normal gate is 0.5 V, which is smaller than the threshold voltage difference between the barrier gate and the normal gate of the device in the main text Figure 3d (1.8V). This is expected because the device reported here has a barrier with lower Gallium content, *i.e.* the built-in tunnel barrier is lower in this case. (c) Color plot of current I as a function of bias voltage $V_{bias}$ and gate voltage $V_{gate}$ (barrier gate and normal gate connected and act as a single local gate). (d) Representative WKB-model-fits of experimental I-V traces at four different values of gate voltage $V_{gate}$. Square-shaped barrier potential is used for the WKB-model-fits. (e) The barrier height as a function of gate voltage, $V_{gate}$. The effective barrier height at 0 V corresponds to the actual conduction band offset between the InSb and $Ga_{0.15}In_{0.85}Sb$ nanowire segments and equals 75 meV. (f) The barrier width as a function of gate voltage, $V_{gate}$. The extracted barrier width is between 10 nm and 20 nm, which agrees with the value extracted from TEM/EDX.